# All Spin Nano-magnetic State Elements

*Sasikanth Manipatruni, Dmitri E. Nikonov, and Ian A. Young*

Exploratory Integrated Circuits, Intel Components Research, Intel Corp

Hillsboro, OR 97124

We propose an all spin state element to enable all spin state machines using spin currents and nanomagnets. We demonstrate via numerical simulations the operation of a state element a critical building block for synchronous, sequential logic computation. The numerical models encompass Landau-Lifshitz-Gilbert (LLG) nanomagnet dynamics with stochastic models and vector spin-transport in metallic magnetic and non-magnetic channels. Combined with all spin combinatorial logic, the state elements can enable synchronous and asynchronous computing elements.

**AUTHOR EMAIL ADDRESS: sasikanth.manipatruni@intel.com**

**CORRESPONDING AUTHOR FOOTNOTE:**

[*]To whom correspondence should be addressed. E-mail: sasikanth.manipatruni@intel.com

As charge based computing approaches the nano-meter scale [1-4], the search for an ultimate switching device for computing is of high importance to ensure the continued scaling in computational efficiency [5]. Several computational state variables including spin [6-8], nanomagnetism [9, 10], orbitronic [11] effects are being explored to enable ultra-low power, normally off and instantly on logic applications. In particular, the spin and magnetic state variables have attracted special attention due to the potential for non-volatile logic and the collective nature of nano-magnetic effects [12, 13]. The spin and magnetic state variables also represent new opportunities for material & interface nano-science [e.g 14-16] and a variety of nano-scale phenomenon [e.g 17-19] which may enable scalable computing for future. However, the spin based computing devices proposed and explored so far lack a critical component for computation, a state element (SE), that enables modern synchronous logic used in microprocessors [20]. In this letter, we propose an all spin state element comprising of nano-magnets interacting via non-local spin currents. We also describe an all spin finite state machine which can be the basis for spin based efficient computing engines.

In this letter, we numerically and theoretically propose a spin state element (SSE) supported by self-consistent nanomagnet and spin transport simulations. The spin transport simulation utilizes vector spin current transport described by 4X4 spin conduction matrices coupled with the dynamics of the nanomagnets. We comprehend the effects of nanomagnetism via a coupled magnetic dynamics simulation with stochastic effects observed at the nano-scale. We also describe an all spin de-multiplexer (SDM) formed with SSE and an spin state machine (SSM) that employs all spin combinatorial logic elements as well as all spin state elements.

We first provide the arguments for the necessity of an all spin state element (SSE) to perform spin based computing. The necessity of an all spin state element can be described as follows. Most modern microprocessors are essentially comprised of state machines (e.g arithmetic logic units (ALUs)) which are controlled by the instruction sets that switch the state machine (e.g ALU) to the appropriate operating mode [21]. Therefore, adopting a new computational state variable [6-11] depends on not only developing combinatorial logic elements but also developing the sequential logic elements such as state machines. Central to a state machine is a collection of controlled state elements which store the state of the machine as decided by the control and input variables. However, for efficient operation of the state machine, it is essential to avoid the repeated state variable conversion. Hence, an all spin state element and all spin state machines that operate using spin currents and nanomagnets at all inputs, outputs and control signals are required.

We describe an all spin state machine (SSM) and a general functional diagram of a spin state state machine in Figure 1. The proposed spin state machine comprises of general purpose combinatorial spin logic (for example implemented as majority gate logic) and spin state elements. The state of the machine ($S_i$) is stored in the all spin state elements of the state register. The present state of the machine ($S_i$) and the present inputs ($\Sigma_i$) are processed in the combinatorial logic unit T which performs the mapping $S_i = T(S_{i-1}, \Sigma)$. The state register stores the output of the logic unit T to the state register comprising of all spin state elements. The all spin state elements are controlled by the output of the clock or the logic for state transition. The output combinatorial block, G calculates the output variables following the mapping $\Lambda_i = G(S_{i-1}, \Sigma)$. Figure 1 B, describes a block diagram of an all spin state element where the nature of the input and output signal is identified. For an all spin state machine it is essential to have an all spin state element which uses both a spin control signal as well as spin input and outputs. Figure 1 C shows a truth table for the operation of the device. The truth table for a state element cannot be implemented using combinatorial logic alone.

The proposed all spin state element comprises of three free magnetic layers and a fixed magnet interacting via spin currents and spin torque exerted by the spin currents [22]. The structure is described in figure 2. The control input magnet (FM1) accepts a spin current injected from a previous all spin logic state [8]. The state element enables or disables the flow of spin current in the channel connecting the magnets M2 and M3. The fixed magnet M0 provides a fixed reference to the magnet M1 to control the state of the SSE. A spin scramble layer (SC) sets up a pure voltage at the intermediate node to provide balanced operation of the device for various operating conditions.

The operation of the device can be described as a non-inverting gate enabled and disabled via a spin current controlled resistor (SCCR). When a spin current comprising of majority electrons with spin parallel/anti-parallel with the fixed magnet (M0) is injected the control magnet (M1), the control magnet responds due to the injected spin torque. The dynamics of the spin torque along with the stochastic nature of the response is described later. The magnets M0 and M1 form a spin current controlled resistor which will be used to control the supply voltage to the non-inverting gate.

The operation of the non-inverting gate for negative supply voltages (inverting gate for positive supply voltages) can be understood as a lateral spin valve with asymmetric spin conditions at the input and output gate [8]. The asymmetric spin conditions can be setup as a result of asymmetric ground conditions or asymmetric spin polarization of the interface or asymmetric areas of overlap with the channels. The non-inverting nature of the gate for a negative supply voltage applied at (Node 3) can be explained as follows: the magnets inject electrons into the channel (from the magnets to the channel, causing a positive current flow from the channel into the magnet). The dominant magnet (e.g FM2 with a larger area of overlap with the channel) sets up a stronger spin voltage (proportional to the spin population density) in the channel under the magnet. A spin current flows from high spin population to low spin population causing a spin current injection into the slave magnet FM3. The inverting nature of the gate can be explained in a similar fashion. However, in the inverting operation, the magnets setup a spin population opposite to their orientation. We will describe the spin current transport dynamics quantitatively later in the article.

We describe an example physical structure of a spin state element comprising of 3D stacked magnetic and non-magnetic metal layers. The example physical device is show in figure 3 where the SCR is formed in the device layer 1 and the controlled non-inverting spin valve is formed in device layer 2. The proposed structure allows for 3D integration for potential scaling path in computational through put per unit area. The device layer 1, comprises of a fixed magnet, spacer (oxide or metal), free magnet assembly as shown in figure 3. The fixed magnet is created using a multilayer combination formed by a suitable anti-ferro-magnet (AFM), synthetic AF stack (with appropriate spacer for exchange coupling control). A typical layer stacking is (IrMn/CoFe/Ru/CoFeB/Spacer/CoFeB). The free magnet is shown at the bottom of the stack. Other variants are possible depending on the process flow and integration issues associated with a back end of line process. A short spin flip length material is formed to create the spin scramble layer. The role of the spin scramble layer is to remove any spin polarization of the current applied to the non-inverting gate formed in the device layer 2.

An example physical structure for the controlled non-inverting gate with the input (FM2) and output magnets (FM3) are shown in figure 3. The non-inverting gate is formed with two free layer magnets (FM2 and FM3) sharing a metallic spin channel (Metal C for e.g made of Cu). The overlap area of the magnets with the channel sets the direction of the information transport. The magnet with the lower overlap with the channel (FM3) acts as the output magnet controlled by the condition of the input magnet. The ground conductance formed by Metal B is chosen such that the current, voltage and energy-delay targets for the device can be optimized.

We now describe the equivalent spin circuit for the spin state element device using a vector spin circuit theory [6]. Vector spin circuit theory models the generation, transport and detection of spin currents using transport models for metallic spin conduction. The equivalent 4X4 conduction elements for the Ferro-magnet (FM) –NM (Normal magnet), spin conduction channel and the SCCR are described in the supplementary material. Figure 5 shows the vector equivalent circuit for a SSE formed with SCCR and a spin logic non-inverting gate. The vector spin conductances $G_{FM0}$, $G_{FM1}$, $G_{FM2}$, and GFM3 describe the spin conduction through the magnets; the spin conduction conductances $G_{SeT}$, $G_{sfT}$ describe the conduction through the spin channel. We represent the metallic spin channel between node 1 and 2 as a combination of two T-equivalent circuits. The conductance $G_{sfT}$ models the loss of spin current due to spin flip in the channel. The $G_{sfT}$ is such that no charge current flows into the ground (open circuit for charge currents) creating a virtual ground for spin currents only. The nanomagnet conductance is modeled by the spin conductance tensors $G_{FM1}(\hat{m}_1)$, $G_{FM2}(\hat{m}_2)$, $G_{FM3}(\hat{m}_3)$. The spin conductance tensors for the nanomagnets vary according to the vector position of the magnet as decided by the nanomagnet dynamics.

We modeled the combined dynamics of the nanomagnets in the presence of spin transport currents through the channel and the SCCR. We first describe a coupled spin transport-magnetization dynamics model for solving the transient dynamics of the SSE comprehending the vector spin flow and the nanomagnet dynamics. The phenomenological equation describing the dynamics of nanomagnet with a magnetic moment unit vector ($\hat{m}$), the modified Landau-Lifshitz-Gilbert-Slonczewski (LLG) equation [22], (see Table 1 for parameters)

$$\frac{\partial \hat{m}}{\partial t} = -\gamma\mu_0 [\hat{m} \times \vec{H}_{eff}(T)] + \alpha \left[ \hat{m} \times \frac{\partial m}{\partial t} \right] + \frac{\hat{m} \times (\vec{I}_s \times \hat{m})}{eN_s} \quad (1)$$

where $\gamma$ is the electron gyromagnetic ratio; $\mu_0$ is the free space permeability; $\vec{H}_{eff}(T)$ is the effective magnetic field due to material/geometric/surface anisotropy, with the thermal noise component (23); $\alpha$ is the Gilbert damping of the material and $\vec{I}_\perp$ is the component of vector spin current perpendicular to the magnetization ($\hat{m}$) leaving the nanomagnet, $N_s$ is the total number of Bohr magnetons per magnet. Implicit in the LLG equation is the fact that absolute values of the magnetic moments of single domain nanomagnets remain constant. The thermal noise plays a critical role in the dynamics of the magnets and manifests as fluctuations to the internal anisotropic field (23). See supplementary material for a description of the noise properties and numerical methods used for stochastic LLG equations. The dynamics of the SSE are solved self consistently with the spin transport in the equivalent circuit models. The LLG solvers pass the condition of the magnets to the spin circuit and the spin circuit solver passed the spin vector current to the LLG solver at each pass of the self-consistent loop till a solution is reached.

The proposed SSE achieves the logic operation of a state element using minimum number of free layer magnetic elements interacting via spin currents. In figure 5, 6 we show the full range of possible states of the SSE. In figure 5 we show the state where the SSE is disabled and the lateral spin valve does not change its state as the supply voltage is varied. In figure 6, we show the state the where the lateral spin valve responds according to the supply voltage following the logic function of the LSV. To verify the operation of the SSE according to the truth table (1C), we apply a varying supply voltage under a fixed input magnet condition (FM2 is fixed at $m_x = -1$ or logic 0).

**SSE state 1: Disabled state of the SSE:**

The simulation of SSE confirms the functionality of the SSE in disabled state under a typical design condition. Both the input and output magnets are effectively disabled using the high impedance state of the controlled resistor. We use the spin scramble layer for providing a scalar (non-spin) voltage as supply voltage to the spin logic unit. The disabled state of SSE can be created by applying a control signal antiparallel to the permanent magnet of the controlled resistor. The disabled state of the SSE is shown in figure 5C where a control signal is applied to the FM1 to set the FM1 magnet in an anti-parallel state to PM. The high resistance state of the PM-FM1 combination disables the lateral spin valve producing the state retention essential for the operation of an SSE. The LSV (and hence the magnets FM2 and FM3) do not respond to the changing supply (which should produce a toggling of the LSV state) or to the input signals at FM2. Figure 5A shows the state of the input (FM2) and output (FM3) magnets where the output magnet is decoupled from any logic state changes. The spin and charge voltages at the node 3, is shown in in figure 6B. The effect of the spin scrambling layer can be seen to remove the residual spin voltage at node 3 close to zero. This is essential in order to obtain a balanced operation of the LSV. The effective charge voltage at the node 3 is below the required value for the operation of the LSV in the disabled state. The Precessional dynamics of the nanomagnets are shown in figure 6 D. The thermodynamic accuracy of the nanomagnet stochastic response is validated using standard models. (See Supplementary information). The nanomagnet precession is initiated by the thermal dynamics inherent to the room temperature nanomagnets.

**SSE state 2: Enabled state of the SSE:**

The simulation of SSE also confirms the functionality of the SSE in enabled state under a typical design condition. Under the enabled state, the output magnet responds to the inputs (or the supply voltage changes) according to the state transition truth table (1C). To verify the operation of the SSE according to the truth table (1C), we applied a varying supply voltage under a fixed input magnet condition (FM2 is fixed at $m_x = -1$ or logic 0). The nanomagnets Precessional dynamics can be seen in figure 7D. The output magnet follows the expected dynamics of an in plane nanomagnet responding to a LSV. The input magnet FM2 is held steady at its original condition verifying the directionality of the LSV (i.e output magnet responds to the input magnet and not viceversa). Figure 7A shows the nanomagnet dynamics of FM2 and FM3 under the enabled condition. The state of the input magnet is constant (as required by directionality of logic) where as the output responds according to the truth table 1C of an enabled SSE. When the supply voltage is positive and the control signal is ON (i.e FM1 parallel to PM), the LSV operates as an inverting gate and the output magnet responds to change its position to logic 1 ($\hat{m}_3=\hat{x}$) at 1 ns. When the supply voltage is negative and the control signal is ON, (i.e FM1 parallel to PM), the LSV operates as a non-inverting gate and the output magnet responds to change its position to logic 1 ($\hat{m}_3=-\hat{x}$) at 1 ns. Hence, the SSE fulfills the truth table shown in figure 1C acting as a inverting/non-inverting gate controlled by the control signal applied to FM1.

In conclusion, we describe an all spin state element employing spin currents and nanomagnets. An all spin state element is a critical enabling component to build a state machine which is the building blocks for all modern computing elements. Combined with all spin combinatorial logic [7, 8], the state elements enable synchronous computing elements. The concept of the SSE can be readily expanded to build all spin flip flops (edge triggered spin state elements), all spin multiplexing and de-multiplexing elements, all spin encoder and decoder circuits for spin computing elements. Supported by recent integration advances [30], all spin state elements and state machines could enable all spin computing elements that complement/augment advanced CMOS technology to enable future always connected, normally off, instantly on computers.

.

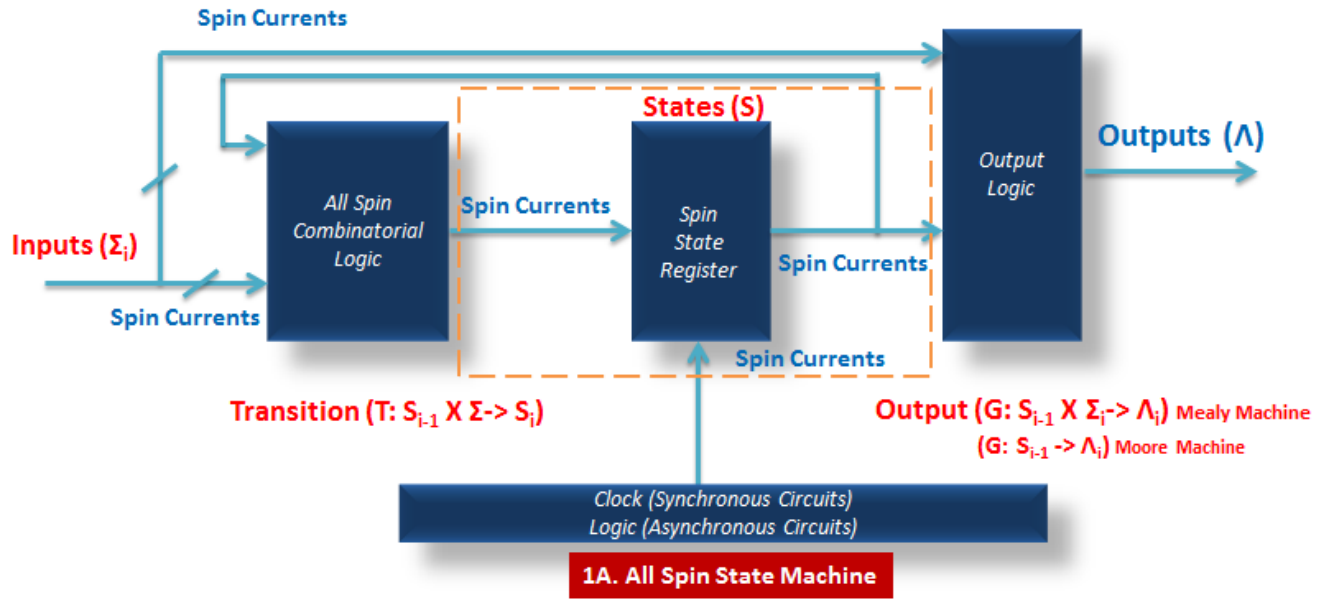

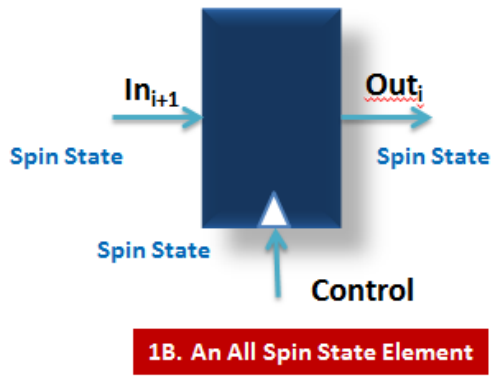

Figure 1 a) An All Spin State Machine comprising of all spin combinatorial logic units and all spin state elements b) An All Spin State Element which is controlled by a spin variable and has spin input and outputs c) Truth table for a state element.

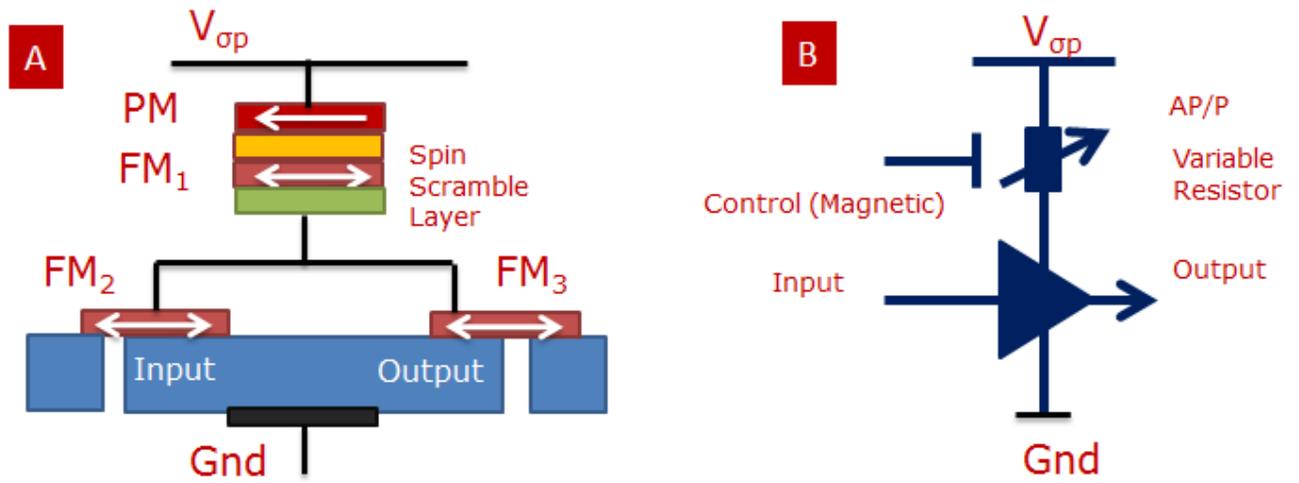

Figure 2 a) Schematic diagram of an All Spin State Element comprising of three free layer nanomagnets and a fixed nanomagnet. The free nanomagnets form a non-inverting gate that is enabled/disabled by the PM+FM1 assembly. B) Equivalent representation of a SSE as a non-inverting gate controlled by a variable resistor.

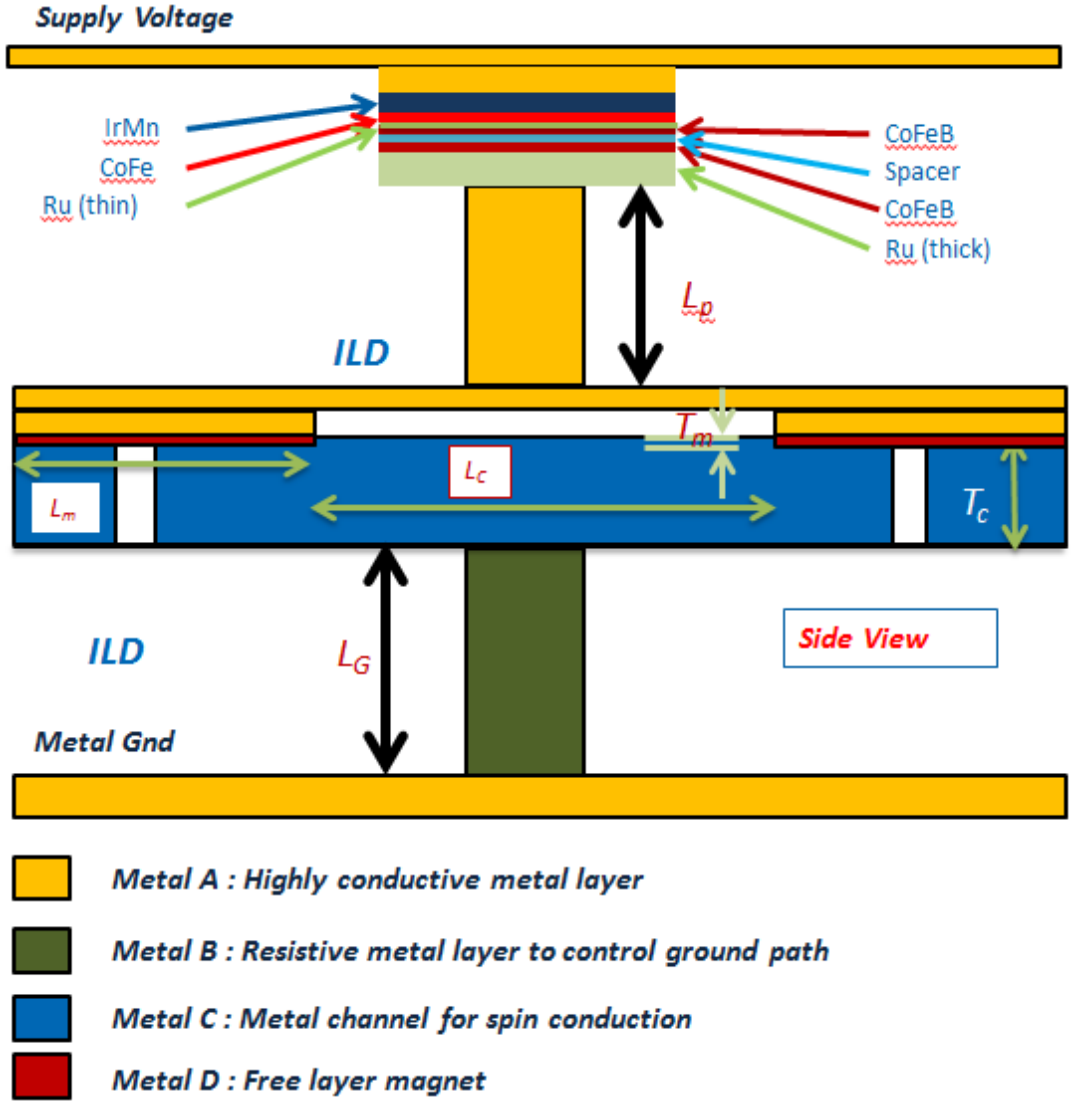

Figure 3 A possible realization of a SSE in a stacked metal layer structure. All the free layer magnets are identical. Spaces between the structures are filled with appropriate interlayer dielectrics.

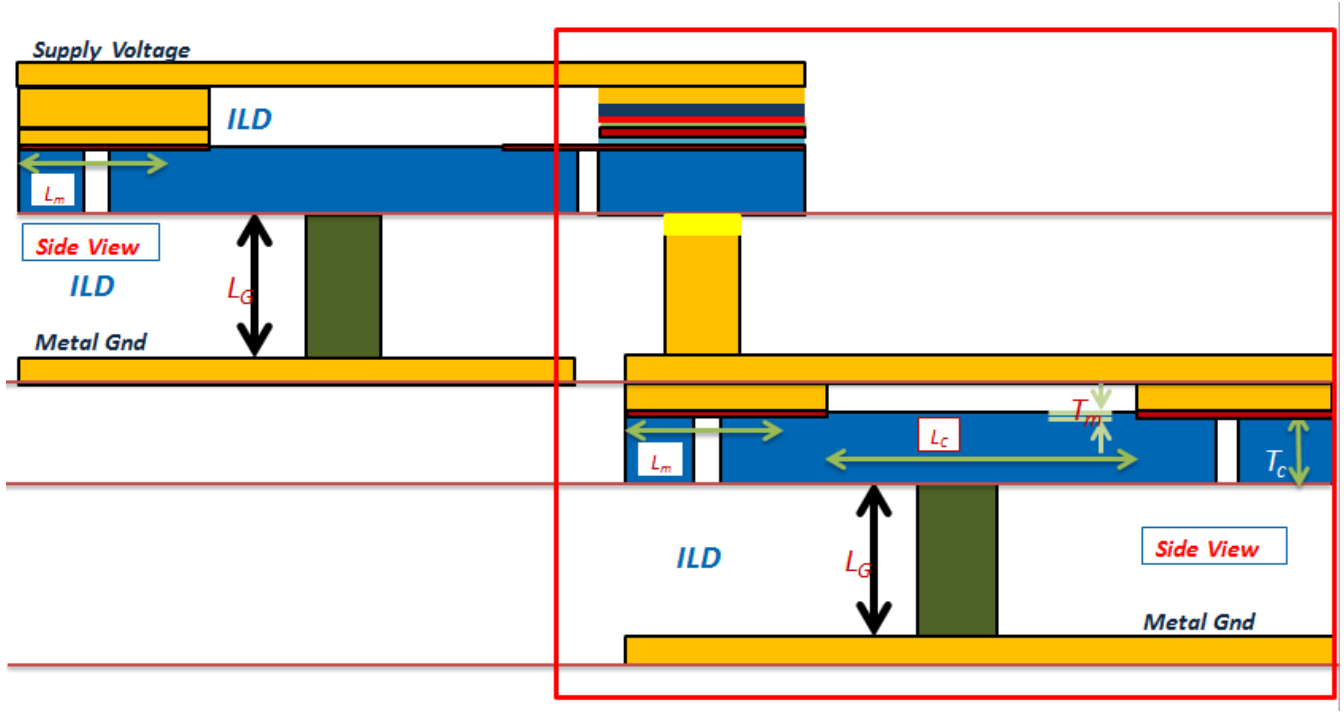

Figure 4 A possible connectivity of a SSE in a stacked metal layer structure where a logic block drives the control magnet. All the free layer magnets are identical. The all SSE scheme is amenable to 3D integration where a state element is distributed across several 3D layers

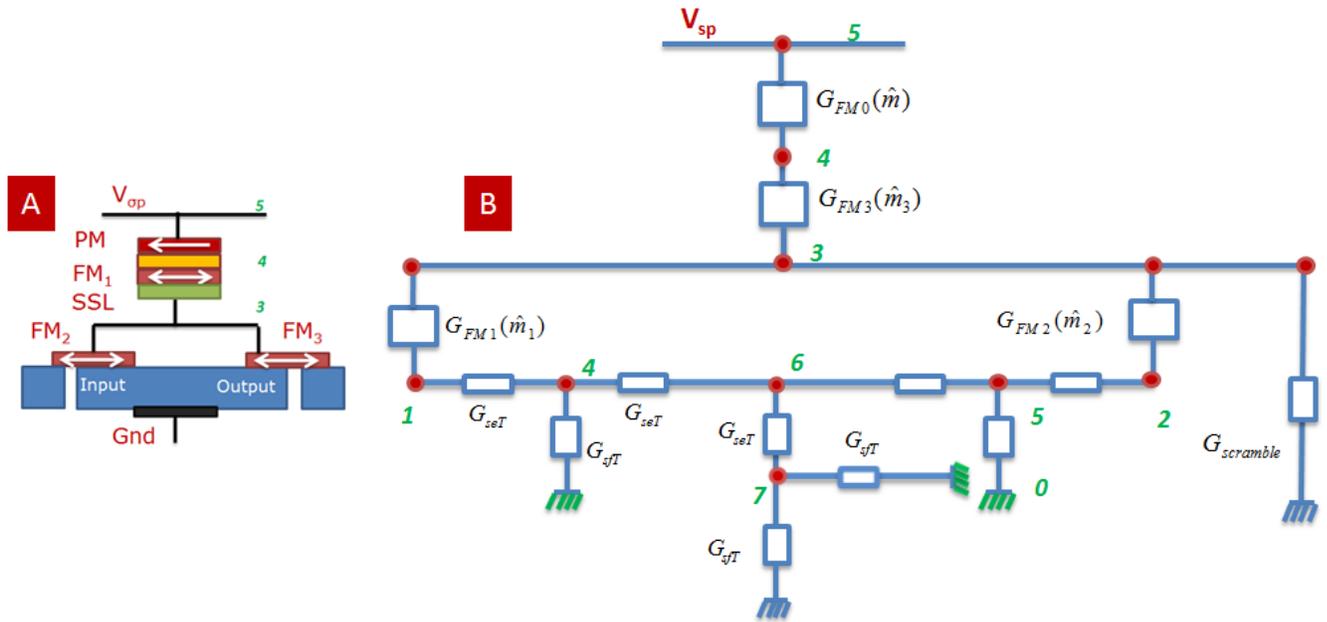

Figure 5: Vector Spin conduction equivalent circuit for SSE. A) SSE with the nodes identified B) the 4 component equivalent circuit model for the SSE.

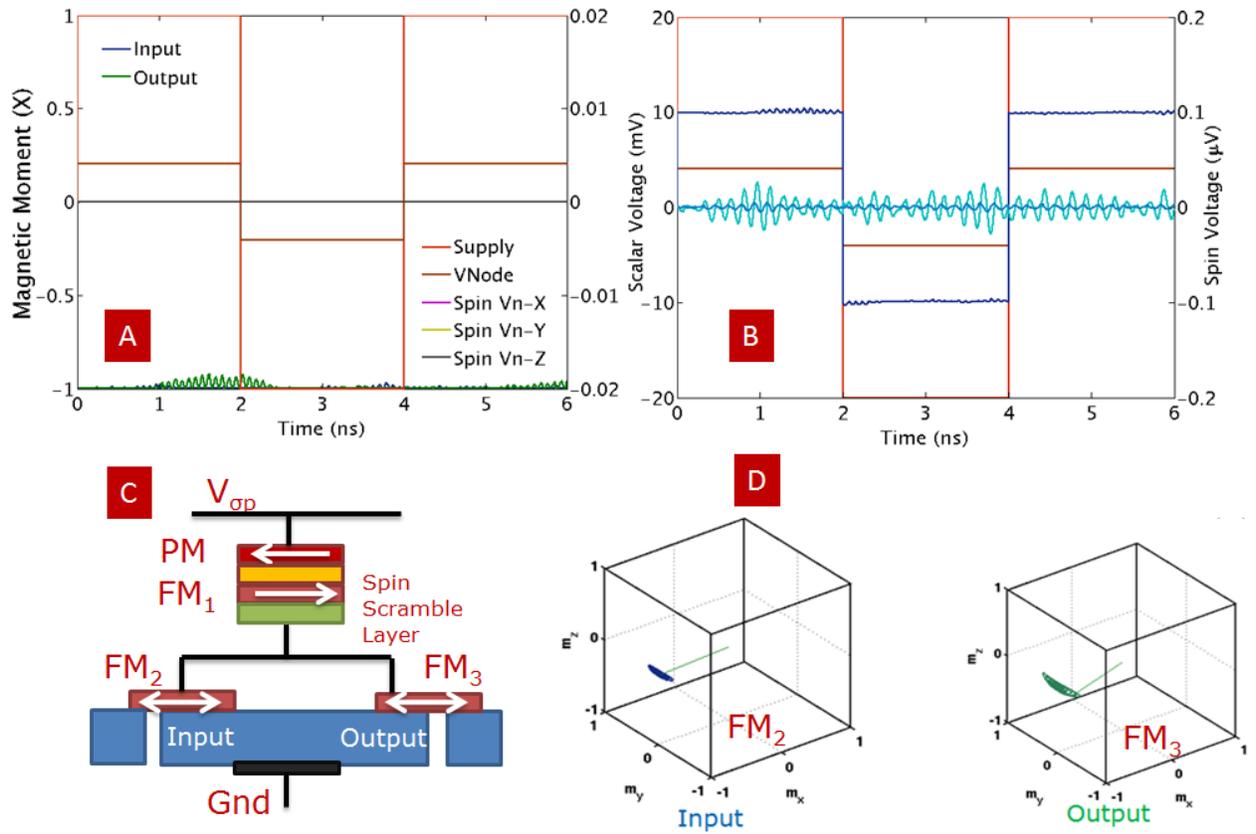

Figure 5: Disabled state of SSE showing the state preservation of the output for varying logic condition.

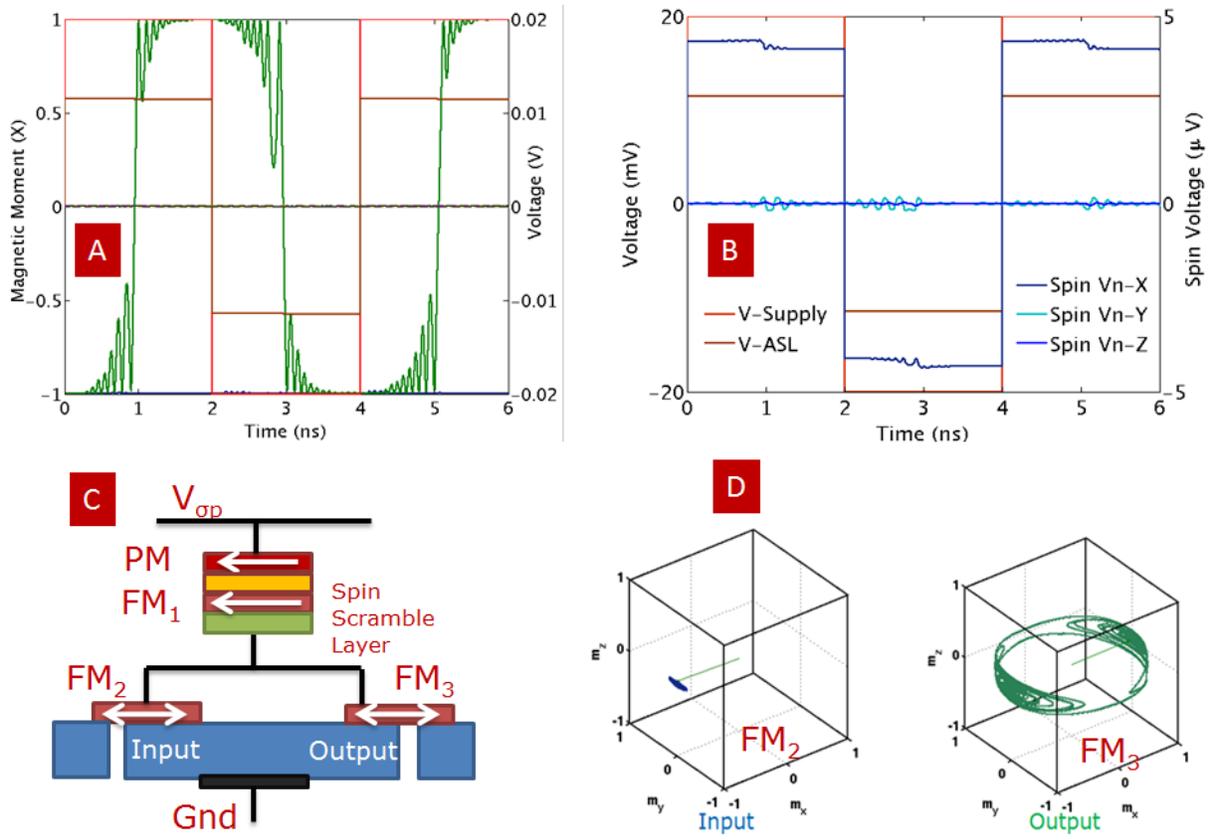

Figure 6: Disabled state of SSE showing the state preservation of the output for varying logic condition.

| Table 1. Nanomagnet parameters for LLG | | | |
|---|---|---|---|
| **Variable** | **Notation** | **Value/Typical Value** | **Units (SI)** |
| Free Space Permeability | $\mu_0$ | $4\pi \times 10^{-7}$ | $JA^{-2}m^{-1}$ |
| Gyromagnetic ratio | $\gamma$ | $17.6 \times 10^{10}$ | $s^{-1}T^{-1}$ |
| Saturation Magnetization of the Magnet | $M_s$ | $10^6$ [24] | A/m |
| Damping of the Magnet | $\alpha$ | 0.007-0.01 [25-27] | - |
| Barrier Height | $E_b$ | 40-100 [28] | kT |
| Effective Internal Anisotropic Field | $H_{eff}$ | $10^3$-$10^6$ [29] | A/m |
| Number of Bohr magnetons in the nanomagnet | $N_s$ | $10^3$-$10^6$ | - |

| Table 2. Parameters used for SSE circuit simulation | | | |
|---|---|---|---|
| **Variable** | **Notation** | **Value** | **Units (SI)** |
| Saturation Magnetization of the Magnet | $M_s$ | $10^6$ | A/m |
| Damping of the Magnet | $\alpha$ | 0.007 | - |
| Effective Internal Anisotropic Field | $H_{eff}$ | $3.06 \times 10^4$ | A/m |
| **Barrier of the magnet** | $\Delta/kT$ | 40 | |
| Length of Magnet | $N_s$ | $10^3 - 10^6$ | - |
| Thickness of Magnet | $T_m$ | 3 | nm |
| Width of Magnet | $W_m$ | 37.8 | nm |
| Length of Magnet | $L_m$ | 75.7 | nm |
| Length of channel | $L_c$ | 100 | nm |
| Thickness of channel | $T_c$ | 200 | nm |
| Length of ground lead | $L_g$ | 200 | nm |
| Thickness of ground lead | $T_g$ | 100 | nm |
| Channel conductivity | $\rho$ | $7 \times 10^{-9}$ | $\Omega \cdot m$ |
| Sharvin conductivity | $G_{sh}$ | $0.5 \times 10^{15}$ | $\Omega^{-1} \cdot m^{-2}$ |
| Polarization | $\alpha_c$ | 0.8 | |